\documentclass{article}
\usepackage{spconf,amsmath,graphicx}
\usepackage{float}
\usepackage{array}
\usepackage{multirow}
\usepackage{adjustbox}
\usepackage{graphicx}
\usepackage{tabularx}
\usepackage{makecell}
\usepackage{adjustbox}
\usepackage{array}
\usepackage{tabularx}
\usepackage{subcaption}

\usepackage{multirow}
\usepackage{booktabs}
\usepackage[pagebackref=false,breaklinks=true,colorlinks,bookmarks=false]{hyperref}
\usepackage{enumitem}
\setlist{nosep, leftmargin=14pt}

\usepackage{mwe} 

\usepackage{lipsum}

\usepackage[table]{xcolor}
\usepackage{colortbl}
\usepackage{graphicx}

\definecolor{pastelblue}{RGB}{192,228,236}   
\definecolor{pastelorange}{RGB}{255,223,186}  
\definecolor{pastelgreen}{RGB}{188,238,188}   
\definecolor{pastelviolet}{HTML}{CCC0DA}

\usepackage{soul}

\title{SynthFM: Training Modality-agnostic Foundation Models for Medical Image Segmentation without Real Medical Data}
\name{Sourya Sengupta$^{1,2*\dagger}$,  Satrajit Chakrabarty$^{1*}$, Keerthi Sravan Ravi$^{1}$, Gopal Avinash$^{1}$, Ravi Soni$^{1}$
\thanks{$^*$ Equally contributing first authors}
\thanks{$^\dagger$ This work was done during the author's internship at GE HealthCare.}
}

\address{$^{1}$GE HealthCare, San Ramon, CA, USA \\
$^{2}$University of Illinois Urbana–Champaign, Urbana,
IL, USA \\
}

\begin{document}
%
\maketitle
\begin{abstract}
Foundation models like the Segment Anything Model (SAM) excel in zero-shot segmentation for natural images but struggle with medical image segmentation due to differences in texture, contrast, and noise. Annotating medical images is costly and requires domain expertise, limiting large-scale annotated data availability. To address this, we propose SynthFM, a synthetic data generation framework that mimics the complexities of medical images, enabling foundation models to adapt without real medical data. Using SAM’s pretrained encoder and training the decoder from scratch on SynthFM’s dataset, we evaluated our method on 11 anatomical structures across 9 datasets (CT, MRI, and Ultrasound). SynthFM outperformed zero-shot baselines like SAM and MedSAM, achieving superior results under different prompt settings and on out-of-distribution datasets.
\end{abstract}
\vspace{-1em}
\section{Introduction}\label{sec:intro}
\vspace{-1em}
Foundation models trained on large datasets, like Meta’s Segment Anything Model (SAM) \cite{ref:sam1} 
 and SAM 2 \cite{ref:sam2}, demonstrate excellent zero-shot segmentation for natural images. Accurate segmentation is vital in computer-aided diagnosis for isolating anatomical structures, but obtaining annotations for medical images is costly and requires clinical expertise. Zero-shot models like SAM offer potential by reducing dependency on task-specific training data. However, SAM performs poorly on medical images due to domain differences; medical images differ from natural images in contrast, texture, and noise, with unclear boundaries in modalities like Ultrasound posing significant challenges.  Therefore, we hypothesize that a zero-shot promptable model like SAM can be adapted for medical image segmentation by training it on synthetic data that closely approximates the data manifold of real-world medical images. In this work, we propose SynthFM, an analytical method of synthetic data generation that captures key data characteristics such as contrast, noise, and textures commonly observed in medical imaging. SynthFM was trained exclusively on this synthetic dataset. We conducted extensive experiments to evaluate its performance on 11 different anatomical structures from three imaging modalities across nine publicly available datasets. The key contributions of this work are:
\begin{itemize}
    \item To the best of our knowledge, SynthFM is the first analytical method to generate synthetic data for training a foundational model to enable generalization to medical imaging. Unlike other foundation models for medical image segmentation, this is the first attempt to create a `foundational data' modeling approach that inherently captures the intrinsic characteristics of medical images.
    \item SynthFM outperforms the original SAM, SAM 2 and UnSAM~\cite{ref:unsam} across datasets, regardless of the number of prompts used.
    \item The model is modality-agnostic and operates in a zero-shot manner, without relying on real-world data, making it suitable for broader applications beyond medical imaging. 
\end{itemize}
\vspace{-1em}
\section{Related Work}\label{sec:format}
\vspace{-1em}
In recent studies, SAM has been explored and adapted in 2D medical image segmentation.  The original SAM model was found to underperform in some medical image segmentation tasks \cite{ref:medsam}. MedSAM \cite{ref:medsam} fine-tuned the original SAM model on a large number of medical images in a supervised manner.  Some other recent works incorporated domain-specific medical knowledge \cite{zhang2023towards} for improving the segmentation performance. 
A more detailed review of SAM in medical image segmentation can be found in \cite{zhang2023towards}.  All of these methods rely on high-quality annotated data, either for fine-tuning or training from scratch on large medical datasets. However, they still face limitations in generalization and scalability due to the dependency on costly, labor-intensive annotations. This highlights a critical gap: the need for a zero-shot approach that eliminates the reliance on annotated data while maintaining strong segmentation performance in medical imaging. 
\vspace{-1em}
\section{Methodology}
\vspace{-1em}
\subsection{Data Generation Strategy}
\begin{figure}[h!]
    \centering
    \begin{subfigure}[b]{\linewidth}
        \centering
        \includegraphics[width=\linewidth, trim=1.3cm 0cm 0.6cm 0cm, clip]{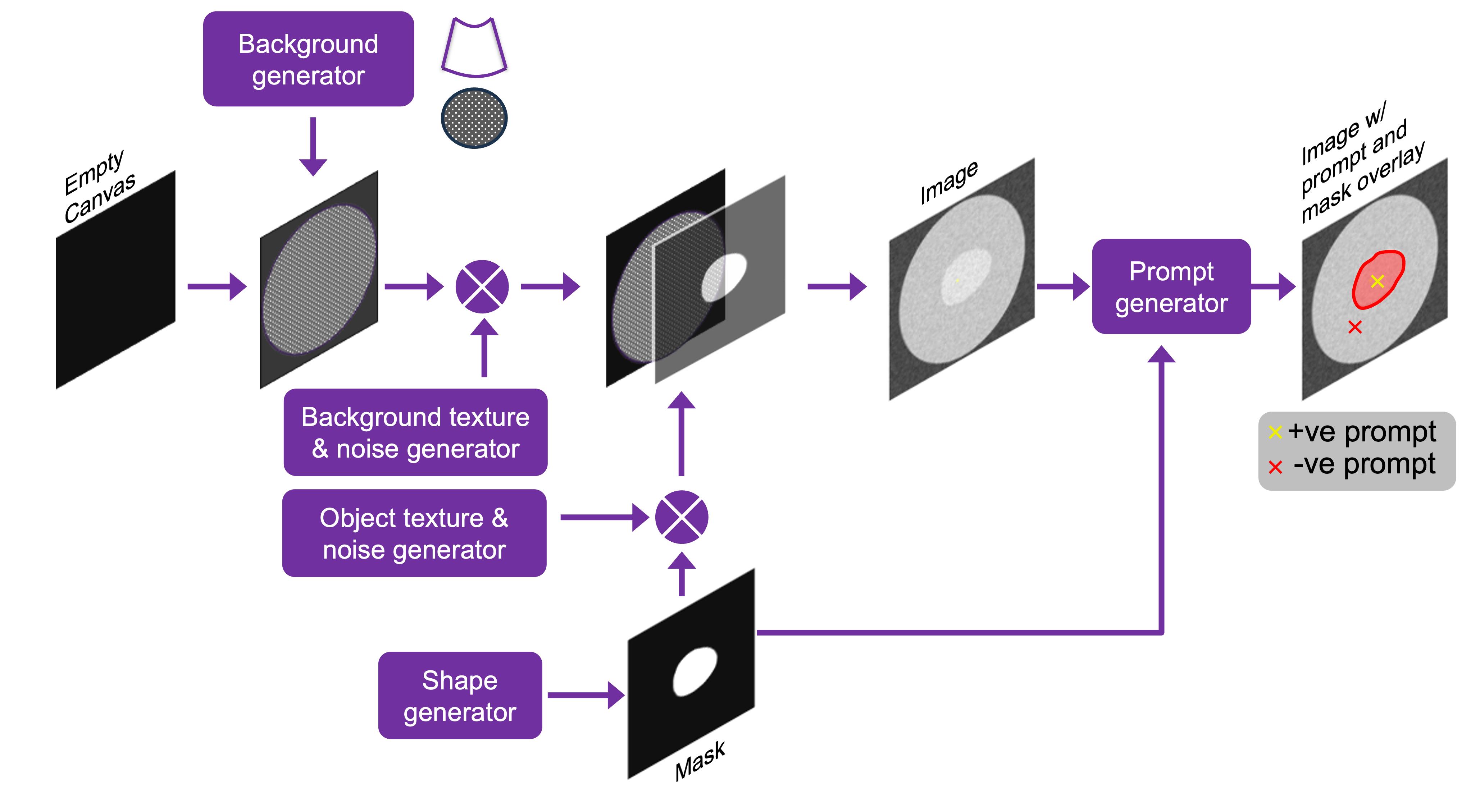}
        \caption{}
        \label{fig:training_method}
    \end{subfigure}

    \begin{subfigure}[b]{\linewidth}
        \centering
        \includegraphics[width=\linewidth, trim=0cm 0cm 0cm 0.2cm, clip]{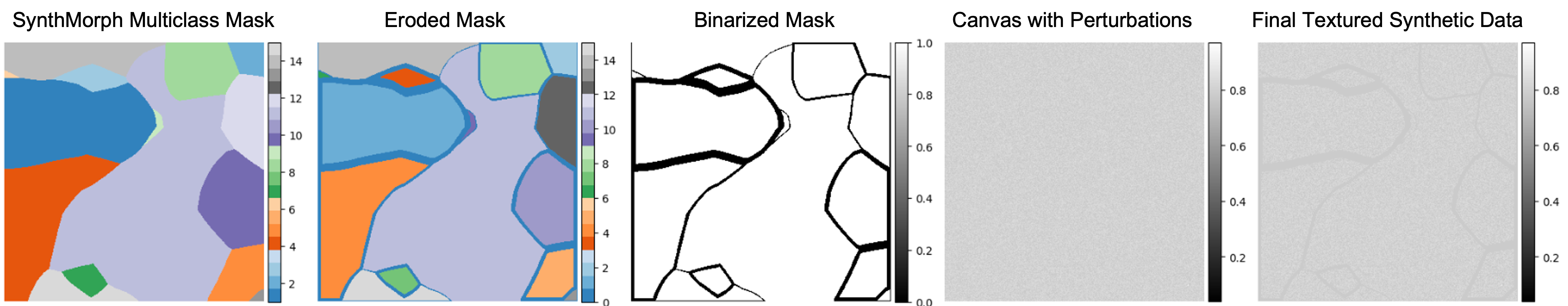}
        \caption{}
        \label{fig:synthmorph_step_by_step}
    \end{subfigure}
    
    \begin{subfigure}[c]{\linewidth}
        \centering
        \includegraphics[width=\linewidth]{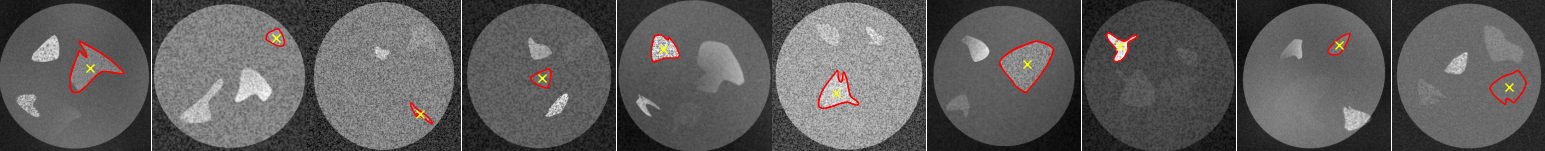}
        \includegraphics[width=\linewidth]{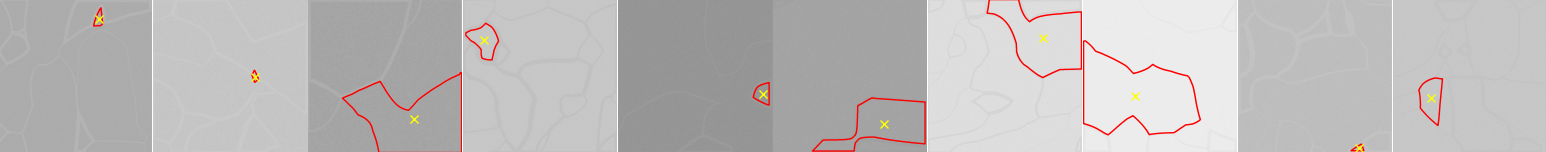}
        \caption{}
        \label{fig:synthetic_data_bezier}
    \end{subfigure}
    \vspace{-1em} 
    \caption{Different stages of data generation of (a) shape-aware and (b) boundary-aware module. (c) Examples of synthetic data generated using shape-aware module (\textit{top}) and boundary-aware module (\textit{bottom}).}
    \vspace{-1em} 
    \label{fig:training}
\end{figure}

\vspace{-0.5em}
SynthFM captures the key aspects of medical imaging, which are different from natural images, such as simulating diverse anatomical shapes, varying contrasts and textures, and generating structures in close proximity to adjacent anatomical features. Figure~\ref{fig:training} illustrates our step-by-step process for generating medically realistic simulated images and examples of generated images. The individual components of SynthFM are described as follows.

\vspace{-1.5em}
\subsubsection{Shape-aware Module} 
\vspace{-0.5em}
Medical images exhibit diverse anatomical shapes and structures, varying across organs and patients. To capture this variability, we adopted a Bézier curve-based object generation approach. A Bézier curve is defined by \( n+1 \) control points \( P_0, P_1, \ldots, P_n \). Its curve \( B(t) \) for \( t \in [0, 1] \) is expressed as \( B(t) = \sum_{i=0}^{n} \binom{n}{i} (1 - t)^{n-i} t^i P_i \), where \( \binom{n}{i} \) is the binomial coefficient. The terms \((1 - t)^{n-i} t^i\) denote Bernstein polynomials, which ensure that the curve starts at \( P_0 \) when \( t = 0 \) and ends at \( P_n \) when \( t = 1 \). The control points heavily influence the curve's shape, pulling it toward these points without necessarily passing through them (except for the endpoints \( P_0 \) and \( P_n \)). By randomly sampling different number of control points, a diverse set of shapes and structures were created in each case.

\noindent\textbf{Texture and Noise Library.}
The unique texture and contrast characteristics of medical images set them apart from natural images, where object-background contrast is typically high. Medical images often feature subtle contrast variations, making structure differentiation more challenging. In our method, objects generated by the shape-aware module were superimposed on a background that was either a black canvas or a black canvas with a phantom (simple circular shape) overlaid, mimicking common imaging modalities, where objects appear against uniform backgrounds.

To simulate the range of noise and texture variations found in real medical images, we developed a noise library, including noise models such as additive Gaussian noise, multiplicative Poisson noise, Perlin noise, speckle noise, and Rician noise. These noise patterns were generated on-the-fly and randomly applied to the shapes during training. Additionally, Gaussian blurring was introduced to replicate the blurriness often seen around organ edges in medical images.

To simulate varying contrast levels between anatomical structures and backgrounds, we modulated the contrast using the heuristic function $(1 - p) \cdot (m \cdot r)$, where \(p\) is the phantom or background intensity, \(m\) indexes each shape (from 1 to the total number of shapes), and \(r\) is a randomly sampled variable from the interval \([-0.2, 0.2]\). The intensity \(p\) was drawn from a uniform distribution \(p \sim \mathcal{U}(0, 1)\) for each case, ensuring a wide range of intensity backgrounds. This formulation generated diverse intensity variations in synthetic images by adjusting contrast between structures and background.
\vspace{-1.5em}
\subsubsection{Boundary-aware Module} 
\vspace{-0.5em}
The shape-aware module captured individual shape complexities but generated shapes independently, resulting in disconnected structures. In medical images, however, structures often share boundaries with minimal contrast differences, complicating segmentation. To address this, we developed the boundary-aware module to generate adjacent structures with shared boundaries.
We leveraged SynthMorph \cite{ref:synth} to generate a multiclass label map with 10-15 labeled clusters. A subset of clusters was then randomly selected and eroded for a randomly chosen number of iterations to introduce boundaries between adjacent structures. After erosion, all clusters were assigned the foreground label and the boundary created from erosion was assigned the background label, resulting in a simplified binary mask.\\ 
\textbf{Contrast and Texture.} 
The training image was generated based on the binary mask by assigning intensity values from a randomly generated canvas with controlled variations. The canvas was populated with values from a randomly chosen range \([limit1, limit2]\), \(limit1 \in [0.1, 0.9]\) and \(limit2\) being either slightly below or above \(limit1\), and within \([0, 1]\). To introduce variability, a small proportion of canvas values was perturbed outside this range, scattered randomly across the entire interval \([0, 1]\). The canvas was then combined with the binary mask: for foreground areas, the values from the random canvas were directly applied. For background areas, the canvas values were averaged and a small random perturbation was added, producing a slight positive or negative shift. This ensured nuanced variations across the image, enhancing the complexity and realism of the synthetic data.

\vspace{-1em}
\subsection{Model architecture and Training details} 
\vspace{-0.5em}
We utilized the SAM architecture in our experiments. Dice loss was used for training, with encoders initialized from the original SAM weights while the decoder was trained from scratch. Each epoch included 10,000 images of size 1024x1024, with a batch size of 1, and data was generated on-the-fly during training. The model was trained for 100 epochs with a learning rate of \(1 \times 10^{-4}\) using cosine decay, and the model with the best validation performance was selected.

\vspace{-1em}
\subsection{Prompt strategy} 
\vspace{-0.5em}
We evaluated SynthFM’s performance using four positive-negative prompt configurations: $(1, 0)$, $(3, 0)$, $(1, 2)$, and $(3, 2)$, where the two values represent the number of positive and negative clicks respectively. For positive prompts, the first prompt was placed near the target structure’s centroid, with additional prompts randomly positioned within the target structure. For negative prompts, the prompts were selected from a dilated region around the target structure—ensuring the first prompt was maximally distant from the centroid, with subsequent negative prompts spaced apart from each other.

\begin{table*}[htbp]
\setlength{\tabcolsep}{2pt}
\renewcommand{\arraystretch}{0.8}
\centering
\caption{Quantitative results of baseline and proposed methods. Highest DSC values per prompt group are highlighted: 
\sethlcolor{pastelblue}\hl{$(1, 0)$}, 
\sethlcolor{pastelorange}\hl{$(1, 2)$}, 
\sethlcolor{pastelgreen}\hl{$(3, 0)$}, 
\sethlcolor{pastelviolet}\hl{$(3, 2)$}. SynthFM values are compared to the highest or next highest DSC values across other methods. Significance is shown as \textbf{\underline{xx.xx}} for p \textless{} 0.001, \textbf{xx.xx} for 0.001 \textless{} p \textless{} 0.05, and xx.xx for p \textgreater{} 0.05.}
\vspace{-1em}
\small
\begin{tabular}{c|c|cccc|cccc|cccc|cccc}
\hline
\multirow{2}{*}{Modality} & \multirow{2}{*}{Structure} & \multicolumn{4}{c|}{UnSAM~\cite{ref:unsam}} & \multicolumn{4}{c|}{SAM~\cite{ref:sam1}} & \multicolumn{4}{c|}{SAM 2~\cite{ref:sam2}} & \multicolumn{4}{c}{\textbf{SynthFM (Ours)}} \\
\cline{3-18}
 &  & \scriptsize{$(1, 0)$} & \scriptsize{$(1, 2)$} & \scriptsize{$(3, 0)$} & \scriptsize{$(3, 2)$} & \scriptsize{$(1, 0)$} & \scriptsize{$(1, 2)$} & \scriptsize{$(3, 0)$} & \scriptsize{$(3, 2)$} & \scriptsize{$(1, 0)$} & \scriptsize{$(1, 2)$} & \scriptsize{$(3, 0)$} & \scriptsize{$(3, 2)$} & \scriptsize{$(1, 0)$} & \scriptsize{$(1, 2)$} & \scriptsize{$(3, 0)$} & \scriptsize{$(3, 2)$} \\
\hline
\multirow{9}{*}{CT} 
 & Aorta & 9.48 & 17.22 & 9.34 & 18.48 & 60.16 & 77.34 & 65.90 & 76.57 & 58.42 & 80.32 & 68.07 & 78.67 & \cellcolor{pastelblue}{\textbf{\underline{72.12}}} & \cellcolor{pastelgreen}{\textbf{\underline{84.47}}} & \cellcolor{pastelorange}{\textbf{\underline{78.05}}} & \cellcolor{pastelviolet}{\textbf{\underline{84.55}}} \\
 & Gallbladder & 3.53 & 11.01 & 3.31 & 14.74 & 26.56 & 56.16 & 35.76 & 54.24 & 8.43 & 63.40 & 24.31 & 65.03 & \cellcolor{pastelblue}{\textbf{\underline{52.18}}} & \cellcolor{pastelgreen}{\textbf{\underline{80.96}}} & \cellcolor{pastelorange}{\textbf{\underline{61.08}}} & \cellcolor{pastelviolet}{\textbf{\underline{80.52}}} \\
 & Kidney Left & 27.46 & 37.48 & 27.99 & 39.31 & 80.72 & 88.55 & 84.28 & 88.93 & 61.44 & 87.37 & 78.75 & 88.68 & \cellcolor{pastelblue}{\textbf{83.78}} & \cellcolor{pastelgreen}{88.79} & \cellcolor{pastelorange}{\textbf{87.51}} & \cellcolor{pastelviolet}{90.29} \\
 & Kidney Right & 23.47 & 30.60 & 27.41 & 38.34 & 75.34 & 86.78 & 78.60 & 86.09 & 56.43 & 84.93 & 73.29 & 86.15 & \cellcolor{pastelblue}{\textbf{\underline{84.00}}} & \cellcolor{pastelgreen}{\textbf{89.50}} & \cellcolor{pastelorange}{\textbf{\underline{87.94}}} & \cellcolor{pastelviolet}{\textbf{\underline{90.26}}} \\
 & Liver & 40.17 & 50.80 & 40.22 & 52.04 & 49.74 & 78.74 & 54.19 & 74.98 & 45.45 & 78.87 & 48.24 & 72.72 & \cellcolor{pastelblue}{\textbf{\underline{73.83}}} & \cellcolor{pastelgreen}{\textbf{\underline{88.36}}} & \cellcolor{pastelorange}{\textbf{\underline{80.03}}} & \cellcolor{pastelviolet}{\textbf{\underline{87.24}}} \\
 & Pancreas & 5.39 & 15.63 & 5.46 & 17.74 & 10.92 & 46.79 & 15.43 & 43.72 & 6.56 & 41.49 & 13.02 & 42.62 & \cellcolor{pastelblue}{\textbf{\underline{43.21}}} & \cellcolor{pastelgreen}{\textbf{\underline{67.38}}} & \cellcolor{pastelorange}{\textbf{\underline{51.55}}} & \cellcolor{pastelviolet}{\textbf{\underline{68.97}}} \\
 & Prostate & 3.79 & 16.57 & 3.79 & 17.69 & 12.33 & 41.10 & 18.07 & 37.97 & 17.06 & 60.01 & 34.53 & 59.66 & \cellcolor{pastelblue}{\textbf{\underline{34.72}}} & \cellcolor{pastelgreen}{\textbf{\underline{78.75}}} & \cellcolor{pastelorange}{\textbf{\underline{54.57}}} & \cellcolor{pastelviolet}{\textbf{\underline{77.42}}} \\
 & Spleen & 12.23 & 22.31 & 12.63 & 24.68 & 37.86 & 70.32 & 48.86 & 66.48 & 20.02 & 64.42 & 32.33 & 63.79 & \cellcolor{pastelblue}{\textbf{\underline{75.59}}} & \cellcolor{pastelgreen}{\textbf{\underline{88.06}}} & \cellcolor{pastelorange}{\textbf{\underline{79.97}}} & \cellcolor{pastelviolet}{\textbf{\underline{87.22}}} \\
\hline
\multirow{9}{*}{MR} 
 & Aorta & 15.64 & 24.05 & 15.13 & 27.84 & 58.87 & 69.24 & 62.17 & 69.19 & 65.35 & 77.00 & 70.69 & 77.47 & \cellcolor{pastelblue}{67.43} & \cellcolor{pastelgreen}{\textbf{\underline{80.15}}} & \cellcolor{pastelorange}{\textbf{74.33}} & \cellcolor{pastelviolet}{\textbf{\underline{80.51}}} \\
 & Gallbladder & 6.87 & 15.68 & 7.62 & 17.96 & 43.49 & 60.40 & 50.39 & 60.14 & 53.82 & 72.31 & 64.69 & 74.65 & \cellcolor{pastelblue}{\textbf{\underline{64.48}}} & \cellcolor{pastelgreen}{\textbf{77.62}} & \cellcolor{pastelorange}{68.08} & \cellcolor{pastelviolet}{77.27} \\
 & Kidney Left & 15.53 & 28.52 & 15.86 & 29.51 & 74.49 & 82.22 & 79.10 & 83.72 & 65.61 & 76.48 & 81.59 & \cellcolor{pastelviolet}{87.48} & \cellcolor{pastelblue}{\textbf{79.78}} & \cellcolor{pastelgreen}{\textbf{85.86}} & \cellcolor{pastelorange}{83.17} & 86.65 \\
 & Kidney Right & 13.03 & 21.75 & 12.99 & 26.74 & 74.37 & 81.80 & 77.85 & 82.62 & 65.22 & 79.30 & 80.54 & 86.13 & \cellcolor{pastelblue}{\textbf{\underline{81.39}}} & \cellcolor{pastelgreen}{\textbf{86.02}} & \cellcolor{pastelorange}{82.73} & \cellcolor{pastelviolet}{86.80} \\
 & Liver & 43.75 & 56.94 & 43.70 & 57.56 & 57.98 & 78.00 & 65.67 & 80.08 & 52.50 & 81.92 & 64.60 & 85.53 & \cellcolor{pastelblue}{\textbf{\underline{71.85}}} & \cellcolor{pastelgreen}{\textbf{\underline{88.54}}} & \cellcolor{pastelorange}{\textbf{\underline{79.99}}} & \cellcolor{pastelviolet}{\textbf{87.99}} \\
 & Pancreas & 5.28 & 13.47 & 5.23 & 13.92 & 17.67 & 46.77 & 26.51 & 46.60 & 21.52 & 48.96 & 33.09 & 55.12 & \cellcolor{pastelblue}{\textbf{\underline{44.53}}} & \cellcolor{pastelgreen}{\textbf{\underline{66.04}}} & \cellcolor{pastelorange}{\textbf{\underline{53.42}}} & \cellcolor{pastelviolet}{\textbf{\underline{67.78}}} \\
 & Prostate & 16.60 & 24.78 & 17.22 & 26.84 & 40.39 & 55.57 & 44.03 & 55.51 & 49.26 & 64.54 & 56.49 & 63.87 & \cellcolor{pastelblue}{\textbf{59.60}} & \cellcolor{pastelgreen}{\textbf{\underline{77.73}}} & \cellcolor{pastelorange}{64.27} & \cellcolor{pastelviolet}{\textbf{76.09}} \\
 & Spleen & 15.75 & 25.38 & 15.35 & 27.52 & 55.94 & 78.24 & 63.82 & 76.50 & 67.36 & \cellcolor{pastelgreen}{87.77} & 75.31 & \cellcolor{pastelviolet}{88.57} & \cellcolor{pastelblue}{\textbf{\underline{80.64}}} & 86.43 & \cellcolor{pastelorange}{\textbf{\underline{82.99}}} & 86.87 \\
\hline
\multirow{4}{*}{US} 
 & Left Atrium & 17.58 & 26.03 & 17.58 & 31.24 & 18.22 & 35.98 & 20.20 & 30.09 & 17.23 & 26.85 & 17.26 & 25.68 & \cellcolor{pastelblue}{\textbf{\underline{57.06}}} & \cellcolor{pastelgreen}{\textbf{\underline{77.11}}} & \cellcolor{pastelorange}{\textbf{\underline{60.75}}} & \cellcolor{pastelviolet}{\textbf{\underline{76.84}}} \\
 & LV Endometrium & 28.56 & 39.96 & 28.56 & 40.87 & 36.90 & 73.54 & 43.73 & 70.06 & 28.45 & 34.28 & 28.83 & 37.40 & \cellcolor{pastelblue}{\textbf{\underline{63.32}}} & \cellcolor{pastelgreen}{\textbf{\underline{82.79}}} & \cellcolor{pastelorange}{\textbf{\underline{67.44}}} & \cellcolor{pastelviolet}{\textbf{\underline{82.23}}} \\
 & Fetal Head & 46.66 & 59.80 & 47.94 & 61.70 & 48.50 & 62.08 & 60.66 & 75.33 & 50.94 & 68.27 & 56.44 & 71.36 & \cellcolor{pastelblue}{\textbf{\underline{67.55}}} & \cellcolor{pastelgreen}{\textbf{\underline{83.84}}} & \cellcolor{pastelorange}{\textbf{\underline{76.97}}} & \cellcolor{pastelviolet}{\textbf{\underline{83.58}}} \\
\hline
\end{tabular}
\label{table:quant_results}
\end{table*}

\vspace{-1em}
\section{Experiments and Results}
\vspace{-0.5em}
\subsection{Datasets and Evaluation Metrics}
\vspace{-0.5em}
We performed evaluations on nine publicly available datasets: CT and MRI data from AMOS~\cite{ref:amos}, CHAOS \cite{ref:chaos}, TotalSegmentatorV2~\cite{ref:total}, and Ultrasound data from CAMUS~\cite{ref:camus}, HC~\cite{ref:hc}, and FH-PS-AOP~\cite{ref:fhps}. For CT and MR, following abdominal organs were included: aorta, gall bladder, left kidney, right kidney, liver, pancreas, prostate, and spleen. For Ultrasound, all structures available in the datasets were included. All data were preprocessed as described in \cite{ref:medsam}. For all performance evaluations, we used the Dice Similarity Coefficient (DSC). A Student’s paired t-test was conducted to determine if the performance differences between SynthFM and the highest or next-highest DSC values across other methods were statistically significant.

\begin{figure}[!htbp]
    \centering
    \vspace{-1em}
    \includegraphics[trim=10 15 10 10, clip, width=0.9\linewidth]{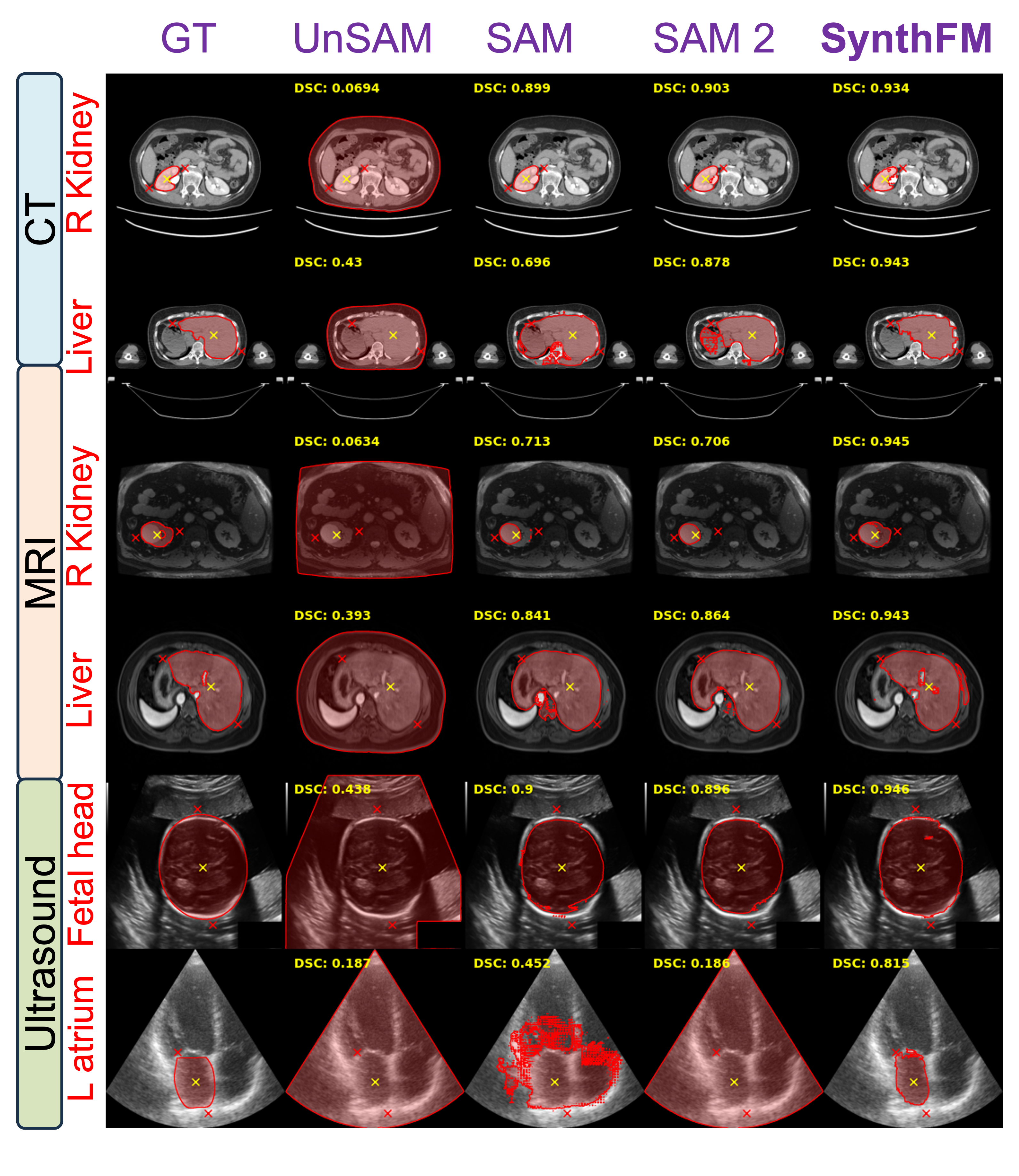}\vspace{-2mm}
    \caption{Qualitative results on different structures across CT, MRI, and Ultrasound modalities for (1 +ve, 2 -ve) clicks. +ve and -ve prompts are shown using \textcolor{yellow}{x} and \textcolor{red}{x}.}
    \vspace{-2em}
    \label{fig:qual_results}
\end{figure}
\vspace{-2em}
\subsection{Comparison with baseline methods}
\vspace{-0.5em}
SynthFM is compared with state-of-the-art segmentation foundation models, including UnSAM~\cite{ref:unsam}, SAM~\cite{ref:sam1}, and SAM 2~\cite{ref:sam2}. The comparisons~(Table~\ref{table:quant_results}, Figure~\ref{fig:qual_results}) highlight the overall superior performance of SynthFM in all three modalities. Especially in Ultrasound, where the contrast is comparatively low and noise is higher, SynthFM outperformed other methods with a significant margin for all organs. In MR, the inherently better contrast of the images narrows the gap between models. Nevertheless, the overall results demonstrate SynthFM's superiority and generalizability across diverse imaging modalities in a complete zero shot setting. Furthermore, across all modalities, increasing the number of positive and negative prompts improves performance for all models.

\vspace{-1em}
\subsection{Ablation Studies}
\vspace{-0.5em}
To evaluate the effectiveness of the individual modules in SynthFM, ablation studies were performed using two variations: (i) with only the shape-aware module and (ii) only the boundary-aware module. Table~\ref{table:ablation_results} reports the mean DSC, averaged across all organs for each imaging modality. The shape module alone performs well on CT and MR, achieving results close to the full model. In contrast, Ultrasound segmentation suffers a notable drop without the boundary module, highlighting its importance for US. Overall, combining both modules yields the best performance across all modalities, underscoring their complementary strengths.
\vspace{-0.5em}
\begin{table}[htbp]
\setlength{\tabcolsep}{3pt}
\renewcommand{\arraystretch}{0.8}
\centering
\caption{Ablation study results for SynthFM modules.}
\vspace{-1em}
\small
\begin{tabular}{c|>{\centering\arraybackslash}p{0.9cm}>{\centering\arraybackslash}p{0.9cm}|>{\centering\arraybackslash}p{0.9cm}>{\centering\arraybackslash}p{0.9cm}|>{\centering\arraybackslash}p{0.9cm}>{\centering\arraybackslash}p{0.9cm}}
\hline
\multirow{2}{*}{Modality} & \multicolumn{2}{c|}{Shape module} & \multicolumn{2}{c|}{Boundary module} & \multicolumn{2}{c}{\textbf{Both}} \\
\cline{2-7}
 & \scriptsize{$(1, 0)$} & \scriptsize{$(3, 2)$} & \scriptsize{$(1, 0)$} & \scriptsize{$(3, 2)$} & \scriptsize{$(1, 0)$} & \scriptsize{$(3, 2)$} \\
\hline
CT & 64.09 & 81.12 & 49.15 & 71.11 & 64.29 & 83.48 \\
\hline
MR & 69.62 & 81.35 & 60.28 & 74.02 & 71.46 & 82.95 \\
\hline
US & 51.37 & 72.38 & 56.11 & 76.89 & 63.86 & 81.55 \\
\hline
\end{tabular}
\label{table:ablation_results}
\end{table}
\vspace{-0.5em}

\subsection{Comparison with Supervised SOTA : MedSAM~\cite{ref:medsam}}
\vspace{-0.5em}
SynthFM is compared to MedSAM \cite{ref:medsam}, the state-of-the-art (SOTA) model for medical image segmentation, trained by fine-tuning the SAM model in a fully supervised manner. To evaluate MedSAM's generalizability, we tested it on ALFI~\cite{antonelli2023alfi}, a label-free microscopy dataset with time-lapse DIC images of HeLa, U2OS, and hTERT RPE-1 cells, a modality unseen by MedSAM during training. MedSAM with click prompts was used for a fair comparison with SynthFM's prompting strategy. The results (Table~\ref{table:compare_sota}) show that SynthFM outperformed MedSAM significantly for interphase and mitosis segmentation. This highlights MedSAM's limited adaptability to unseen modalities and SynthFM's robust generalizability, making it well-suited for diverse imaging scenarios.
\vspace{-0.5em}
\begin{table}[htbp]
\setlength{\tabcolsep}{2pt}
\renewcommand{\arraystretch}{0.8}
\centering
\caption{Quantitative comparison with supervised baseline for four prompt configurations.}
\vspace{-1em}
\small
\begin{tabular}{c|cccc|cccc}
\hline
\multirow{2}{*}{Structure} & \multicolumn{4}{c|}{MedSAM~\cite{ref:medsam}} & \multicolumn{4}{c}{\textbf{SynthFM (Ours)}} \\
\cline{2-9}
 & \scriptsize{$(1, 0)$} & \scriptsize{$(1, 2)$} & \scriptsize{$(3, 0)$} & \scriptsize{$(3, 2)$} & \scriptsize{$(1, 0)$} & \scriptsize{$(1, 2)$} & \scriptsize{$(3, 0)$} & \scriptsize{$(3, 2)$} \\
\hline
Interphase & 12.88 & 12.35 & 13.91 & 11.89 & \cellcolor{pastelblue}40.73 & \cellcolor{pastelorange}55.57 & \cellcolor{pastelgreen}50.23 & \cellcolor{pastelviolet}59.46 \\
Mitosis    & 61.55 & 54.79 & 61.21 & 51.11 & \cellcolor{pastelblue}70.65 & \cellcolor{pastelorange}70.89 & \cellcolor{pastelgreen}68.66 & \cellcolor{pastelviolet}69.45 \\
\hline
\end{tabular}
\label{table:compare_sota}
\end{table}
\vspace{-1em}
\section{Conclusion}
\label{sec:conclusion}
\vspace{-1em}
This is the first work to propose a systematic synthetic data generation framework for interactive, modality-agnostic, promptable segmentation in medical imaging. Trained solely on synthetic data, our model outperformed zero-shot baselines like SAM, SAM 2, and UnSAM across CT, MR, and Ultrasound modalities. Unlike other models, our approach is rooted in a `foundational data' modeling framework. Future work will explore additional prompt settings and extend to 3D medical image segmentation.

\end{document}